\documentclass[12pt]{iopart}

\usepackage{graphicx}
\usepackage{dcolumn}
\usepackage{bm}
\usepackage[cp1250]{inputenc}
\usepackage{ifthen}
\usepackage{iopams}

\newcommand{\bk}{\mathbf{k}}
\newcommand{\bq}{\mathbf{Q}}
\newcommand{\bQ}{\mathbf{Q}}

\newcommand{\eq}{\begin{equation}}
\newcommand{\eqx}{\end{equation}}
\newcommand{\eqn}{\begin{eqnarray}}
\newcommand{\eqnx}{\end{eqnarray}}

\begin{document}

\title[Unconventional superconducting phases in a correlated Fermi gas]{Unconventional superconducting phases in a correlated two-dimensional Fermi gas of nonstandard quasiparticles: a simple model}

\author{Jan Kaczmarczyk$^1$, and Jozef Spałek$^{1, 2}$}

\address{$^1$ Marian Smoluchowski Institute of Physics, Jagiellonian University, ul. Reymonta 4, 30-059 Kraków, Poland}
\address{$^2$ Faculty of Physics and Applied Computer Science, AGH University of Science and Technology, ul. Reymonta 19, 30-059 Kraków, Poland}

\ead{\mailto{kaczek@gmail.com}, \mailto{ufspalek@if.uj.edu.pl}}

\date{\today}

\begin{abstract}
We discuss a detailed phase diagram and other microscopic characteristics on the applied magnetic field - temperature ($H_a-T$) plane for a simple model of correlated fluid represented by a two-dimensional (2D) gas of heavy quasiparticles with masses dependent on the spin direction and the effective field generated by the electron correlations. The consecutive transitions between the Bardeen-Cooper-Schrieffer (BCS) and the Fulde-Ferrell-Larkin-Ovchinnikov (FFLO) phases are either continuous or discontinuous, depending on the values of $H_a$ and $T$. In the latter case, weak metamagnetic transitions occur at the BCS-FFLO boundary. We single out two different FFLO phases, as well as a reentrant behaviour of one of them at high fields. The results are compared with those for ordinary Landau quasiparticles in order to demonstrate the robustness of the FFLO states against the BCS state for the case with spin-dependent masses (SDM). We believe that the mechanism of FFLO stabilization by SDM is generic: other high-field low-temperature (HFLT) superconducting phases benefit from SDM as well.
\end{abstract}


\submitto{\JPCM}

\maketitle

\section{Introduction}
Unconventional superconductivity in heavy-fermion and organic-metal systems is studied almost as frequently as high-temperature superconductivity and comprises a number of heavy-fermion and organic metallic systems \cite{1}. Among the states observed and discussed intensively recently is the superconductivity in the systems without space \cite{2}, and time \cite{3} inversion symmetry, the Fulde-Ferrell-Larkin-Ovchinnikov (FFLO) state \cite{4} and the states in which magnetic order, usually antiferromagnetic (AF), coexists with the FFLO \cite{AFexp, AFth} or the Bardeen-Cooper-Schrieffer (BCS) type of state. Also, the FFLO states are discussed recently in the context of cold atomic fermionic gases \cite{5} and quark-gluon plasma \cite{6}. One of the basic motivations for these studies is the circumstance that the normal state can be represented by a Fermi fluid, albeit almost localized, so the nature of paired state can be rationalized to a greater detail. Also, the intriguing feature of those superconductors is a cooperation rather than competition with magnetism \cite{7b}.
The field-induced magnetism can be incorporated into the FFLO state, since there are substantial portions of the Brillouin-zone volume over which the quasiparticle excitations are gapless.

It is the later topic (the FFLO appearance) which is the principal subject of this paper starting from a two-dimensional (2D) $d$-wave superconductor composed of unconventional (correlated) quasiparticles. Namely, we represent the heavy-fermion liquid by a gas of quasiparticles with the spin-direction dependent effective masses (SDM), which were indeed observed in CeCoIn$_5$ and other systems \cite{McCollam} (in that case $m^* \equiv m_\sigma$, $\sigma = \pm 1$ being the particle spin quantum number). Another non-trivial feature of our approach is the inclusion of the effective field $h_{cor}$ acting upon the magnetic moments in the spin-polarized state \cite{8}. Both characteristics are generated by the electron correlations treated in the mean-field-type schemes \cite{8,9}. The experimental motivation for our study is the observation of both SDM \cite{McCollam} and FFLO (or FFLO mixed with magnetism) \cite{AFexp} in the same heavy fermion system CeCoIn$_5$, and the question we tackle is whether these two phenomena are interconnected. To address it we consider a simplest situation of electron gas with the FFLO state in the simplest form (FF type with $\Delta(\mathbf{r}) = \Delta_0 e^{i \mathbf{q} \mathbf{r}}$). We also consider a $d$-wave form of the superconducting gap, i.e. $\Delta_{\bk, \bQ} = \Delta_\bq (\cos{k_x} - \cos{k_y})$, where $\bQ$ is the Cooper pair momentum ($\bQ \neq 0$ in FFLO state). Such form of the gap reflects the principal feature of quasi-two-dimensional superconductivity in strongly correlated electrons \cite{10}. We show that the phase diagram with these high-field low-temperature (HFLT) $d$-wave superconducting phases in 2D differs remarkably from its 3D correspondant with the $s$-wave symmetry \cite{11,previous}. Namely, several FFLO states appear in the present situation even when we disregard the possibility of their coexistence with antiferromagnetism \cite{AFexp,AFth}. We also show that a weak metamagnetic transition accompanies the BCS $\rightarrow$ FFLO discontinuous transition.

\section{Model: unconventional gas of quasiparticles with real-space pairing}
The principal features of our approach have been defined earlier \cite{11} (cf. particularly Sections V - VII). We start from the effective quasiparticle picture which is common to both narrow-band and hybridized correlated-electron systems \cite{8,9,13,14}. The explicit form of quasiparticle energies in the gas of correlated quasiparticles is

\begin{equation}
    \xi_{\bk \sigma} = \frac{\hbar^2 k^2}{2 m_\sigma} - \sigma (h + h_{cor})  - \mu, \label{eq:disp}
\end{equation}

where $\mu$ is the chemical potential, $h = g \mu_B H_a$, and the mass enhancement factor in the large-$U$ limit \cite{14} in the simplest situation is

\eqn
    \frac{m_\sigma}{m_B} & = & \frac{1-n_\sigma}{1-n} = \frac{1-n/2}{1-n} - \sigma \frac{\overline{m}}{2(1-n)} \equiv \nonumber \\
                         & \equiv & \frac{1}{m_B} (m_{av} - \sigma \Delta m/2), \label{eq:m}
\eqnx

with $\overline m \equiv n_\uparrow - n_\downarrow$ being the spin polarization and $n = n_\uparrow + n_\downarrow$ the band filling. Also $\sigma = \pm1$ is the spin quantum number, $m_B$ is the band mass, $\Delta m \equiv m_2 - m_1$ is the mass difference and $m_{av} \equiv (m_1 + m_2) / 2$ is the average mass. As one can see, the spin-dependent mass enhancement is particularly strong for an almost half-filled case when $1-n \equiv \delta \ll 1$, i.e. for the quasiparticles close to the Mott-Hubbard localization. Here the superconducting phases in this 2D $d$-wave superconductor are discussed in detail and compared briefly with the previous results \cite{11,previous}. In connection with this one should note that the concept of SDM has been also used in the context of coexistence of ferromagnetism and superconductivity \cite{Ying}.

Even though our considerations represent a model situation, we assume the following values of the parameters, emulating the heavy fermion systems: the filling $n = 0.97$, the elementary square-cell area $S = (4.62 \AA)^2$, the starting ($H_a = 0$) quasiparticle mass $m_{av} = 100 m_0$ (data for CeCoIn$_5$ \cite{McCollam}), the pairing potential cutoff and magnitude $\hbar \omega_C = 17\textrm{ K}$, and $V_0 = 90\textrm{ K}$, respectively. The characteristic energy scale associated with spin-fluctuations in CeCoIn$_5$ is $T_{sf} = 10\textrm{ K}$ \cite{Petrovic} - a value comparable to our $\hbar \omega_C$. For those parameters, the chemical potential was equal to $\mu \approx 126\textrm{ K}$. This means that $V_0 \lesssim \mu$ and the (weak-coupling) BCS approximation can be regarded only as a proper solution on a quantitative level at best. Additionally, the chemical potential is readjusted in the superconducting state so that $n$ is constant. The pairing potential has the separable $d$-wave form 

\eq
V_{\bk, \bk'} = - V_0 (\cos{k_x} - \cos{k_y}) (\cos{k_x'} - \cos{k_y'}),
\eqx

which differs slightly from that used in Ref. \cite{previous}d. For thus defined quasiparticles with energy $\xi_{\bk \sigma}$, we derive their correspondants $E_{\bk \sigma}$ in the superconducting states \cite{11,Shimahara}

\eqn
E_{\bk \sigma} & = & E_\bk + \sigma \xi^{(a)}_\bk, \\
E_\bk & = & \sqrt{\xi^{(s)2}_\bk+\Delta_{\bk, \bq}^2}, \\
\xi^{(s)}_\bk & \equiv & \frac{1}{2} (\xi_{\bk + \bq/2 \uparrow} + \xi_{-\bk + \bq/2 \downarrow}), \\
\xi^{(a)}_\bk & \equiv & \frac{1}{2} (\xi_{\bk + \bq/2 \uparrow} - \xi_{-\bk + \bq/2 \downarrow}),
\eqnx

as well as the free-energy functional $\mathcal{F}$ and the system of four self-consistent integral equations for the field $h_{cor}$, magnetization $\overline{m}$, gap magnitude $\Delta_\bQ$ and the chemical potential $\mu$. Explicitly, starting from the free energy functional $\mathcal{F}$, we obtain the corresponding integral equations of the following form

\begin{eqnarray}
 \mathcal{F} &=& - k_B T \sum_{\bk \sigma} \ln (1 + e^{-\beta E_{\bk \sigma}}) + \sum_\bk (\xi^{(s)}_\bk - E_\bk) + \nonumber \\
 && + N \frac{\Delta_\bQ^2}{V_0} + \mu N + \frac{N}{n} \overline{m} h_{cor}, \label{eq:sc1} \\
 h_{cor} &=& - \frac{n}{N} \sum_{\bk \sigma} f(E_{\bk \sigma}) \frac{\partial E_{\bk \sigma}}{\partial \overline{m}} + \nonumber \\
 && + \frac{n}{N} \sum_\bk \frac{\partial \xi_\bk^{(s)}}{\partial \overline{m}} \Big( 1 - \frac{\xi_\bk^{(s)}}{E_\bk} \Big), \label{eq:sc2} \\
 \overline{m} &=& \frac{n}{N} \sum_{\bk \sigma} \sigma f(E_{\bk \sigma}), \label{eq:sc3}\\
 \Delta_\bq &=& \frac{V_0}{N} \sum_\bk (\cos{k_x} - \cos{k_y})^2 \times \nonumber \\
 && \times \frac{1 - f(E_{\bk \uparrow}) - f(E_{\bk \downarrow})}{2 E_\bk} \Delta_\bq, \label{eq:sc4} \\
 n &=& n_\uparrow + n_\downarrow = \nonumber \\
   &=& \frac{n}{N} \sum_{\bk \sigma} \{u_\bk^2 f(E_{\bk \sigma}) + v_\bk^2 [1 - f(E_{\bk, -\sigma})]\}. \label{eq:sc5}
\end{eqnarray}

Those quantities determine the physical free energy in different (BCS, FFLO, NS) states which are compared to obtain the phase diagram and other microscopic characteristics, as we discuss below. Note that we limit ourselves to a single $\bQ$ (Fulde-Ferrell type) solution \cite{11,Shimahara} as we intend to describe superconductivity with SDM in the simplest case and thus test the importance of the quasiparticles mass spin-direction dependence. In that situation, the whole problem comprises a simultaneous solution of those four integral equations for $\mu$, $\overline{m}$, $\Delta_\bQ$ and $h_{cor}$ for fixed $\bQ$ followed by subsequent minimization of thus obtained physical free energy $\mathcal{F}$ with respect to $\bQ$.

\begin{figure}
\includegraphics[height=16cm]{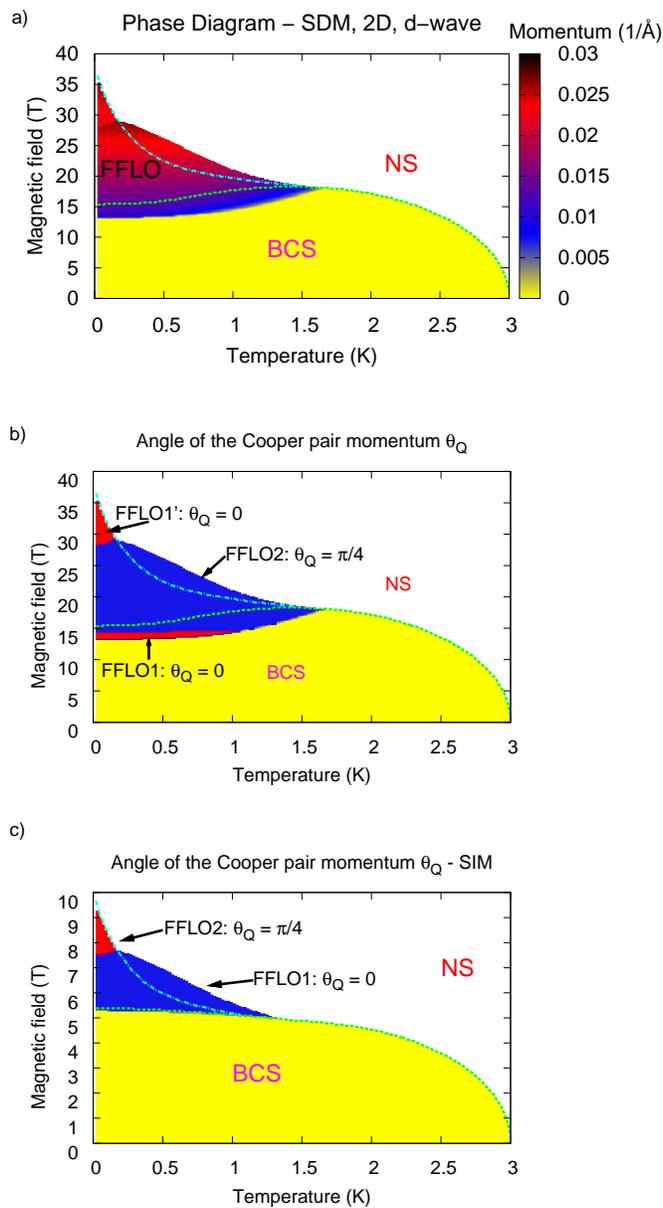}
\caption{\label{fig:PHD} (Colour online). Phase diagram for the cases with the spin-dependent (SDM), a), b) and with spin-independent masses (SIM), c). Light (yellow) regions correspond to $\bq = 0$ (BCS phase), the darker (blue, red) to the state with $\bq \neq 0$ (FFLO phase) and the white to normal state (NS). The colour scale in a) is defined by the pair momentum $\bq$. Note that for SDM with increasing temperature, the transition from BCS to FFLO state occurs at higher fields, in qualitative agreement with experimental results \cite{AFexp}. The different FFLO phases are exhibited in b). The red region corresponds to the Cooper-pair momentum $\bq$ in the $k_x$ direction ($\theta_\bq = 0$), whereas the blue one to the momentum along the diagonal ($k_x = k_y$, $\theta_\bq = \pi/4$). Note that this anisotropy results solely from the $d$-wave gap symmetry, as the unpaired gas is isotropic. The dashed line marks the BCS critical field $H_{c2}$ in the Pauli limit \cite{Hc2}, and the dot-dashed line marks $H_{c2}$ for the solution with $\theta_\bq = 0$.}
\end{figure}

\section{Results: BCS vs FFLO states}

The overall phase diagram in 2D case on the applied magnetic field ($H_a$) - temperature (T) plane is exhibited in figure \ref{fig:PHD} for the cases with spin-dependent (SDM) (a, b) and the spin-independent (SIM) (c) effective masses. The FFLO phase is robust only in the former case, as for the $s$-wave solution for the three-dimensional gas \cite{11}, although the difference is greater in the 3D case. The specific difference is that in the present case two distinct phase-boundary lines appear inside the FFLO state, as detailed in Figure 1b: the topmost and the lowest parts (red colour) have the Cooper-pair momentum $\bQ$ oriented along the $k_x$ (or $k_y$) direction, whereas the middle phase (blue colour) has $\bQ$ along the diagonal ($k_x = k_y$). Also, superconductivity of FFLO type exists up to the field of $35\textrm{ T}$ in the SDM case, i.e. the field more than 4 times larger than that for the SIM case. Hence, the former system \textit{indeed} belongs to the class of \textit{high-field low-temperature superconductors}.

\begin{figure*}
\scalebox{0.7}{\includegraphics[angle=270]{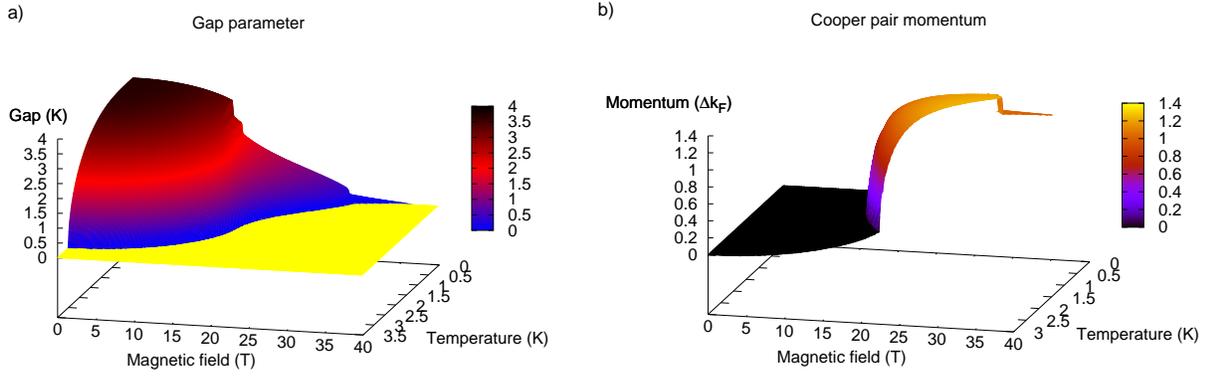}}
\caption{\label{fig:2} (Colour online). a) Gap parameter $\Delta_\bQ$ (in units of K) and b) Cooper pair momentum $\bQ$ in units of the Fermi momentum difference $\Delta k_F \equiv k_{F \uparrow} - k_{F \downarrow}$, both on $H_a$ - $T$ plane. Transitions between various phases are seen as a change of the magnitude of the gap: the lower-field transition are first-order, whereas the transition to normal state is continuous.}
\end{figure*}

To visualize the detailed nature of the transition to the FFLO phase we have plotted in figure \ref{fig:2} profiles of the gap magnitude $\Delta_\bQ$ and the Cooper pair momentum $|\bQ|$, both on the $H_a$ - $T$ plane. In the low-$T$ limit the observed gap jumps meaning that the transitions BCS $\rightarrow$ FFLO1 ($\bQ \parallel k_x \, \textrm{axis}$), as well as the transition FFLO1 $\rightarrow$ FFLO2 ($\bQ \parallel (k_x, k_y) \, \textrm{diagonal}$) and FFLO2 $\rightarrow$ FFLO1' ($\bQ \parallel k_x \, \textrm{axis}$) are discontinuous, whereas the transition to the normal state is continuous (cf. also \cite{9}). As the temperature increases, all the transitions (except that from FFLO2 to FFLO1') become continuous, but the exact position of the terminal bicritical point will not be discussed in detail here. The phase FFLO1' illustrates a reentrant high-field behaviour for FFLO1 phase (cf. also figure \ref{fig:PHD}c for the SIM case). Note also that the FFLO states exist far beyond the second critical field $H_{c2}$ \cite{Hc2} for BCS state, marked by the dashed line.

\begin{figure*}
\scalebox{1.2}{\includegraphics{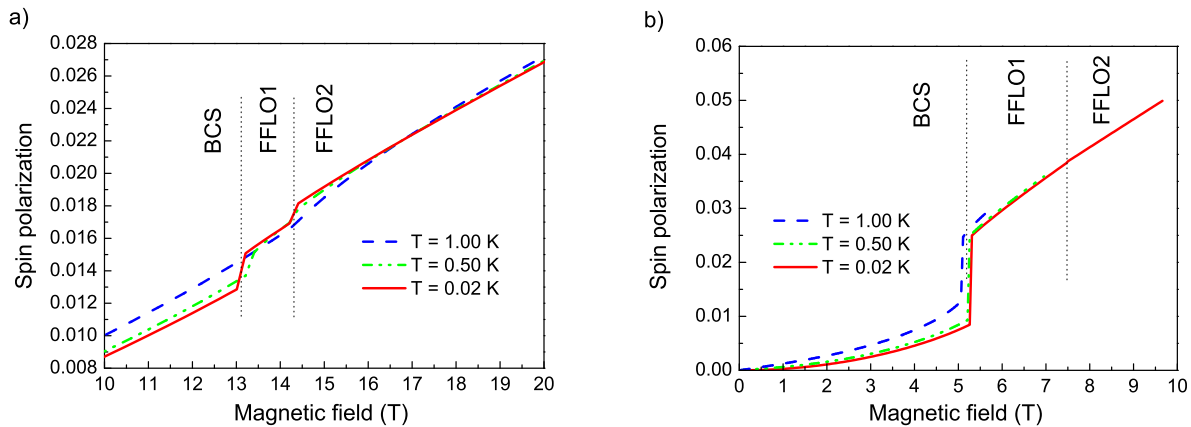}}
\caption{\label{fig:3} (Colour online). Spin polarization $\overline{m} \equiv n_\uparrow - n_\downarrow$ as a function of applied field. Note the weak jumps corresponding to the discontinuous transitions at $T = 0.02\textrm{ K}$ and $T = 0.50\textrm{ K}$ for SDM case (a) and much larger in the SIM case. (b). For the SDM case all transitions at $T = 1\textrm{ K}$ are continuous.
}
\end{figure*}

\begin{figure}
\scalebox{0.33}{\includegraphics[angle=270]{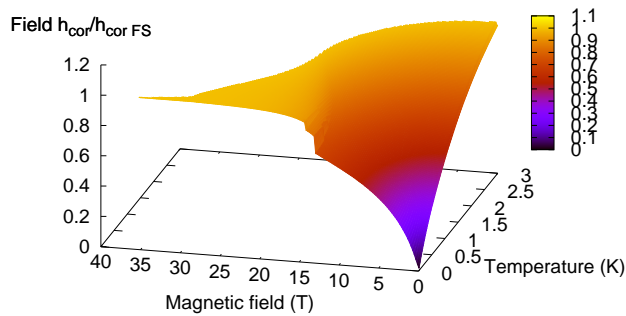}}
\caption{\label{fig:4} (Colour online). Correlation induced effective field $h_{cor}$ in the superconducting state on $H_a-T$ plane relative to that in the normal state $h_{corFS}$, which is typically equal to $-0.33 h$, and therefore the field $h_{cor}$ compensates applied field $H_a$.}
\end{figure}

The above phase transitions can be connected with the magnetization changes. This is because the FFLO phase encompasses semimacroscopic regions of $\bk$-space with gapless quasiparticle excitations in the superconducting phase. This means that the magnetization curve will show a nontypical behaviour, particularly in the vicinity of the transition to FFLO state, as displayed in figure \ref{fig:3}. Namely, the $\overline{m}(H_a)$ exhibits a weak metamagnetic behaviour accompanied by a weak jump at the two lower-field transition points. It is surprising at first look that the corresponding jump is much larger in the SIM case. However, one must remember that in the SDM case the field $h_{cor}$ compensates largely the applied field (see figure \ref{fig:4} for details). The spin magnetization does not include the magnetic dipole moment which may arise from an inhomogeneous current-carrying state when $\bQ \neq 0$. Obviously, the FFLO state may coexist with AF (or SDW) or spin-flop phases, but these cases are not discussed here, as we would like to characterize in detail here the "pure" FFLO state to single out its novel features in the SDM case.

To compare our results with those for three-dimensional system and $s$-wave pairing symmetry we recall here the mechanism behind the FFLO stabilization by SDM presented in \cite{11} (cf. Section VI there). Namely, SDM compensate the Zeeman effect influence by reducing the Fermi wave vectors splitting. Therefore, superconducting state with SDM has higher critical fields (here $h_{c2} = 10\textrm{ T}$ for SIM, and $h_{c2} = 36\textrm{ T}$ for the SDM case, cf. figure \ref{fig:PHD}). The FFLO state benefits from SDM by a greater extent than BCS because spin polarization $\overline{m}$ in the latter is smaller (cf. figure \ref{fig:3}), and from (\ref{eq:m}) the mass difference $\Delta m \propto \overline{m}$. Therefore, in BCS the mass difference is smaller, and the Fermi wave vectors splitting larger than in FFLO (the Zeeman term influence is compensated less effectively). Hence, at $T=0$ the FFLO fills about $1/2$ of the phase diagram for SIM, and about $2/3$ for SDM. On the other hand, as temperature $T$ increases, the spin polarization increases in the BCS state (see figure \ref{fig:3}) allowing larger mass difference $\Delta m$ and reducing Fermi wave vectors splitting enhancing superconductivity. This is why the transition line between BCS and FFLO is curved upwards in the SDM case. In the present situation, the BCS state can have a substantial spin-polarization already at $T=0$ (unlike in the 3D, $s$-wave case) and therefore the BCS state can benefit from SDM already at $T=0$, and the FFLO state is not stabilized so spectacularly here, as it was in the 3D case (where in the BCS phase $\overline{m} \approx 0$ at $T=0$).

For the sake of completeness, we draw in figure \ref{fig:4} the effective field induced by the correlations. The jumps reflect the discontinuous transitions discussed above. The field $h_{cor}$ (in units of $h_{corFS}$ for the unpaired Fermi sea) increases both with increasing temperature and field. The mass difference $\Delta m = m_\downarrow - m_\uparrow$ changes with the applied field reflecting the change in $\overline{m}(H_a)$; the relative difference $\Delta m / m_{av}$ reaches about 10$\%$ for the applied field of the order of $30\textrm{ T}$.

\section{Conclusions}

In summary, we have singled out different FFLO states in a 2D gas of correlated quasiparticles with spin-dependent effective masses (SDM) and effective field induced by the electron correlations, as well as compared them briefly with those in SIM case \cite{previous}. A number of FFLO phases appears and these phases are stable in an unusually high fields only for the case with SDM which were indeed discovered in CeCoIn$_5$ and other systems \cite{McCollam}. It is suggested that these nonstandard properties of quasiparticle states should be their universal feature for all the systems close to the $f$- or $d$-electron Mott-Hubbard localization if the atomic disorder effects are very weak. Namely, other HFLT phases (including various FFLO phases mixed with antiferromagnetism) can be stabilized from having SDM as well, since they always have higher spin-polarization than the uniform superconducting state, and then SDM compensate the Zeeman term influence more effectively than in the uniform superconducting state. Extension of these results to incorporate the antiferromagnetic ordering within the present approach and for realistic highly anisotropic three dimensional electronic structure, would most probably provide a decisive answer about the nature of high-field low-temperature phase in CeCoIn$_5$ \cite{AFexp} and organic systems \cite{Singleton}.

\ack
The work was supported by Ministry of Higher Education and Science, Grants Nos. N N202 128736 and N N202 173735. The project was performed under the auspices of the COST P-16 ESF Grant, entitled \textit{"Emergent Behavior in Correlated Matter"} (ECOM), as well as of the National Network \textit{"Strongly Correlated Systems"}, and the Marie Curie TOK Grant MTDK-CT-2004-517186 (COCOS).

{}


\begin{thebibliography}{99}
\bibitem{Hc2}
    $H_{c2}$ is determined as the applied field in which the BCS solution becomes unstable (its free energy is higher than the normal-state free energy).
\end{thebibliography}
\end{document}